\documentclass[aps,reprint,twocolumn,prl,showpacs,superscriptaddress,floatfix]{revtex4-1}
\usepackage{amssymb,amsmath,amsfonts} 
\usepackage{epsfig,graphicx}
\begin{document}
\title{Anomalous thermalisation in quantum collective models}

\author{Armando Rela\~{n}o}
  \affiliation{Departamento de F\'{\i}sica Aplicada I and GISC,
    Universidad Complutense de Madrid, Av. Complutense s/n, 28040
    Madrid, Spain}
\email{armando.relano@fis.ucm.es}
\begin{abstract}
  We show that apparently thermalised states still store relevant
  amounts of information about their past, information that can be
  tracked by experiments involving non-equilibrium processes. We
  provide a condition for the microcanonical quantum Crook's theorem,
  and we test it by means of numerical experiments. In the
  Lipkin-Meshkov-Glick model, two different procedures leading to the
  same equilibrium states give rise to different statistics of work in
  non-equilibrium processes. In the Dicke model, two different
  trajectories for the same non-equilibrium protocol produce
  different statistics of work. Microcanonical averages provide the
  correct results for the expectation values of physical observables
  in all the cases; the microcanonical quantum Crook's theorem fails,
  in some of them. We conclude that testing quantum fluctuation
  theorems is mandatory to verify if a system is properly thermalised.

\end{abstract}


\maketitle

{\em Introduction.-} The amazing development of experimental
techniques during the last two decades \cite{Bloch:08,Langen:15} has
spurred on the research on the foundations of quantum thermodynamics
\cite{Gemmer:09}. An important number of these techniques deal with
coherent atomic systems, with Hilbert spaces growing not
exponentially, but linearly, with the number of atoms. As paradigmatic
examples, we highlight recent experimental results involving systems
with just one
\cite{Greiner:02,Gross:10,Albiez:05,Trenkwalder:15,Martin:13,Gerving:12,Will:10},
and two semiclassical degrees of freedom \cite{Baumann:10}. Non-usual
thermodynamics have been already reported on some of them
\cite{Alcalde:12,Bastarrachea:16,Perez-Fernandez:17}. In this Letter
we show that the process of equilibration and thermalisation is also
anomalous in the Dicke \cite{Dicke:54} and the Lipkin-Meshkov-Glick
(LMG) \cite{Lipkin:65} models, realized in some of the previously
quoted experiments \cite{Gross:10,Albiez:05,Baumann:10}. Apparently
thermalised states keep relevant amounts of information about their
past, information that can be tracked by experiments dealing with
non-equilibrium protocols.

Consider an isolated quantum system evolving from a pure initial
state, $\left| \psi(0) \right>$. Although the time-evolved state,
$\rho(t)=\left| \psi(t) \right> \left< \psi(t) \right|$, remains
always pure, it does stay close to an effective equilibrium state,
$\overline{\rho} = \lim_{T \rightarrow \infty} (1/T) \int_0^T dt
\rho(t)=\sum_n \left| \left< \psi(0) \right| \left. E_n \right>
\right|^2 \left| E_n \right> \left< E_n \right|$, during the majority
of the time \cite{Cazalilla:10,Polkovnikov:11,Eisert:15,Gogolin:16},
being $\left| E_n \right>$ the eigenstate with energy $E_n$. As a
consequence, the expected values of representative observables
${\mathcal O}$ are well described by $\overline{{\mathcal
    O}}=\text{Tr} \left[ \overline{\rho} {\mathcal O} \right]$
\cite{Neumann:29,Equilibrium}. But in general $\overline{{\mathcal
    O}}$ is different from the microcanonical average, $\left<
  {\mathcal O} \right>_{\text{mic}}={\mathcal O}(E)$; they are similar
only if $\left( \Delta E \right)^2 \left| {\mathcal O}'' \left( E
  \right) / {\mathcal O} \left( E \right) \right| \ll 1$
\cite{Srednicki:96}, where $\left( \Delta E \right)^2 = \sum_n \left|
  C_n \right|^2 \left(E-E_n \right)^2$ measures the energy width of
the initial state. This requirement is usually fulfilled in chaotic
systems, in which the majority of the eigenstates are {\em typical}
\cite{Neumann:29}, and ${\mathcal O} \left( E_n \right)$ is almost
equal to the microcanonical average ${\mathcal O} (E)$ for every
eigenstate around the system energy $E$ ---what is called Eigenstate
Themalisation Hypothesis (ETH) \cite{ETH}.

Proper thermalisation also entails important consquences for
non-equilibrium processes. A number of fluctuation theorems
\cite{Jarzynski:97,Crooks:99,Talkner:07,Campisi:11} state that the
statistics of work only depend on the properties of the initial
equilibrium states.  Let us consider a forward process, $\alpha_i
\rightarrow \alpha_f$, starting from an eigenstate $E$ of an intial
Hamiltonian $H \left( \alpha_i \right)$, and the corresponding
backwards one, $\alpha_f \rightarrow \alpha_i$, starting from an
eigenstate with $E+w$ of a final Hamiltonian $H \left( \alpha_f
\right)$. Under almost any circumstances, the following equality
always holds, independenly of the {\em trajectory} followed by the
protocol \cite{Dalessio:16},
\begin{equation}
\frac{P_f (E,\alpha_i,w)}{P_b (E+w,\alpha_f,-w)} = \frac{g(E+w,\alpha_f)}{g(E,\alpha_i)}.
\label{eq:qft}
\end{equation}
$P_f(E,\alpha_i,w)$ is the probability of investing the work $w$ in
the forward protocol; $P_b(E+w,\alpha_f,-w)$, the probability of
obtaining the same quantity in the backwards; $g(E,\alpha_i)$, the
density of states of the initial Hamiltonian at energy $E$; and
$g(E+w,\alpha_f)$, the one of the final Hamiltonian at energy
$E+w$. If both initial states are {\em exact} microcanonical ensembles,
the same equality holds; this is called the microcanonical quantum
Crook's theorem \cite{Talkner:08,nota1}.

{\em Condition for the microcanonical quantum Crook's theorem.-} Let
us now consider that the same protocol is performed from the
actual equilibrium state, $\overline{\rho}$. Eq. (\ref{eq:qft}) is
only applicable if
\begin{equation}
\begin{split}
\left| g''(E)/g(E) + P''(E)/P(E) + \right. \\ \left. + P'(E) g'(E) / \left[ P(E) g(E) \right] \right| \left( \Delta E \right)^2 \ll 1,
\end{split} 
\label{eq:condition}
\end{equation}
where $g(E) \equiv g(E+w,\alpha_f)$ ($g(E) \equiv g(E,\alpha_i)$) for
the forward (backwards) process, and $P(E)$ is the probability of the
transition from $\left| E \left( \alpha_i \right) \right>$ to $\left|
  E + w \left( \alpha_f \right) \right>$, which is assumed to be a
smooth function of the energy $E$
\cite{Supl}. Eq. (\ref{eq:condition}) depends on both the density of
states and the transition probabilities, so this condition might be
more or less demanding for different trajectories of the same
protocol. 

{\em Different initial states.-} In this part of the Letter we test
the consequences of Eq. (\ref{eq:condition}) on the LMG model
\cite{Lipkin:65}, applicable to a number of physical situations
\cite{Gross:10,Unanyan:03,Micheli:03,Morrison:08,Larson:10}. 
It decribes the dynamics of $N$ two-level atoms, each level
represented by a kind of scalar bosons, $s$ and $t$,
\begin{equation}
H = \alpha t^{\dagger} t - \frac{1-\alpha}{N} Q \cdot Q,
\end{equation}
where $Q=s^{\dagger} t + t^{\dagger} s$; $N$ is the number of atoms
(which is conserved), and $\alpha$ is the only external parameter of the
model. In the thermodynamical limit, $N \rightarrow \infty$, this
model is well described by a semiclassical Hamiltonian with just
one degree of freedom \cite{Vidal:06,ESQPT_LMG}
\begin{equation}
H(q,p) = \alpha p^2 + \left( 5 \alpha -4 \right) p \left(1-p \right) - 4 \left(1 - \alpha \right)p \left(1-p \right) \sin^2 q.
\end{equation}
As a consequence, both the level density, $g(E,\alpha) = \int dq dp
\, \delta\left[E-H(q,p)\right]$, and the microcanonical averages,
$\left< {\mathcal O}(E,\alpha) \right> = \int dq dq \, {\mathcal
  O}(q,p) \delta\left[E-H(q,p)\right]/g(E,\alpha)$, can be
analytically calculated \cite{Supl}.

The consequences of Eq. (\ref{eq:condition}) are tested by preparing
the system at particular values of $E$ and $\alpha$, by means of two different
procedures: {\em (i)} The system is quenched to the final value of the
external parameter, $\alpha_{\text{ini}} \rightarrow \alpha$. {\em
  (ii)} The system is first {\em pre-quenched} to an intermediate
value of the external parameter, $\alpha_{\text{ini}} \rightarrow
\alpha_{\text{int}} \sim \alpha$. Then, it is {\em agitated}
by repeteadly quenching $\alpha_{\text{int}} \rightarrow \alpha$ (and
viceversa), letting it relax after any of these quenches. The
procedure is repeated until the required value of the energy $E$ is
reached, at the final value of $\alpha$.

\begin{figure}[h!]
  \includegraphics[width=0.65\linewidth,angle=-90]{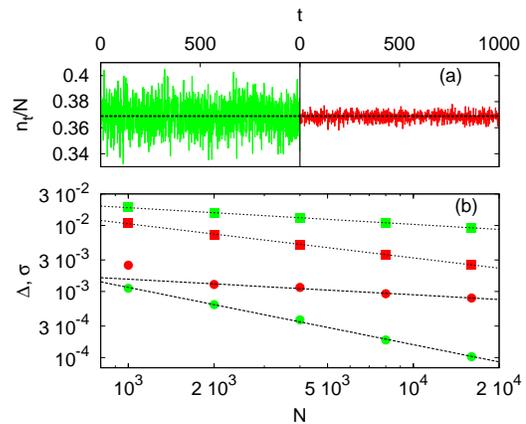}
\caption{{\em Panel (a)}. Expected values for $n_t/N$ in the LMG model, for
  $\alpha=0.5$, $E/N=-0.24$, and $N=1.6 \cdot 10^4$. Left column (green online), state
  prepared following procedure (i); right column (red online), with procedure (ii). Black dotted line, analytical
  expected values, $\left< {\mathcal O}(E,\alpha) \right>$. {\em Panel (b)}. Scaling as a function of
  the system size. Solid circles represent the distance between
  long-time averages and microcanonical expected values,
  $\Delta=\left| \overline{{\mathcal O}}-{\mathcal O}_{\text{mic}}(E,
  \alpha) \right|$. Solid squares, the variance of the
  fluctuations around the equilibrium state, $\sigma^2=(1/T) \int_0^T
  dt \, \left( {\mathcal O}(t) - \overline{{\mathcal
      O}}\right)^2$. Light symbols (green online), results obtained
  following procedure (i); dark symbols (red online), results obtained
  following procedure (ii). Results are double-averaged: over the four
  selected observables, and over $10$
  different sets of points: $\alpha=0.2$ and $E/N=-0.6$ together with
  $9$ different cases for $\alpha=0.5$: from $E/N=-0.26$ to
  $E/N=-0.22$, with steps $\Delta E/N=0.005$. Dotted and dashed lines
  represent fits to power-law behaviors $N^{-\gamma}$. (See main text
  for details).}
\label{fig:termo}
\end{figure}

\begin{figure}[h!]
  \includegraphics[width=0.7\linewidth,angle=-90]{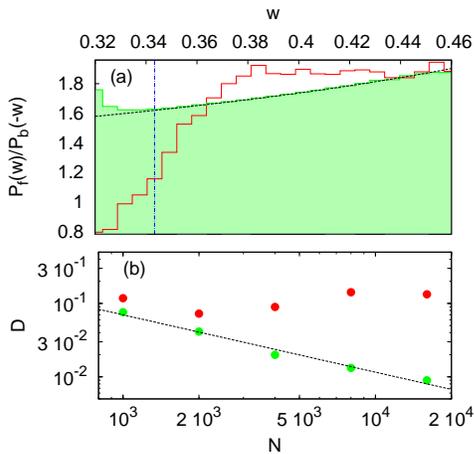} 
  \caption{{\em Panel (a)}. Statistics of work resulting from forward
    $\alpha_i=0.2 \rightarrow \alpha_f=0.5$, and backwards
    $\alpha_i=0.5 \rightarrow \alpha_f=0.2$ processes, in the LMG
    model with $N=1.6 \cdot 10^4$. Filled histogram (green online),
    initial equilibrium states prepared following procedure (i). Empty
    histogram (red online), following procedure (ii). Dotted black
    line, results of
    Eq. (\ref{eq:qft}). Vertical line, particular case studied in
    panel (a) of Fig. \ref{fig:distr}. {\em Panel (b)}. Scaling of the distance, $D=(1/N) \sum_i
    \left| h(w_i) - h_{\text{QFR}}(w_i) \right| / h_{\text{QFR}}
    (w_i)$, where $h(w_i)$ is the height of the actual histogram at
    work $w_i$; $h_{\text{QFR}}$ the theoretical value,
    Eq. (\ref{eq:qft}), and $N$ the number of bins. Light circles
    (green online), results following procedure (i). Dark circles (red
    online), results following procedure (ii). Dashed black line,
    power-law fit, $N^{-\gamma}$.}
\label{fig:qft}
\end{figure}

To test if the system is properly thermalised we focus on four
observables $n_t/N = p$, $n_t^2/N^2=p^2$, $n_t n_s/N^2= p(1-p)$, and
$Q \cdot Q/N^2 = 4p(1-p) \cos^2 q$. In panel (a) of
Fig. \ref{fig:termo} we show long-time dynamics of $n_t/N$, together
with the corresponding microcanonical averages (see caption for
details), for both the procedure (i) (left part), and the procedure
(ii) (right part). In all the cases, $\alpha_{\text{ini}}$ depends on
the target energy. For procedure (ii), $\alpha_{\text{int}}=0.25$ if
$\alpha=0.2$, and $\alpha_{\text{int}}=0.53$ if $\alpha=0.5$
\cite{Supl}.  In panel (b), we present a quantitative test to
corroborate that the system is properly thermalised. We display how
the size of the fluctuations (squares) and the distance between
long-time and microcanonical averages (circles) scale with the system
size. Displayed data are the result of a double averaging: over $10$
different initial states $(E,\alpha)$ (see caption of
Fig. \ref{fig:termo} for details), chosen according to the findings
coming from Fig. \ref{fig:qft}, and over the four selected
observables. The size of the fluctuations around the equilibrium state
decreases with the system size following power laws, $N^{-\gamma}$,
with $\gamma = 0.253(3)$ for procedure (i), and $\gamma=0.517(9)$, for
procedure (ii). This means that the system remains close to the
equilibrium state during the majority of the time. The distance
between microcanonical and long-time averages also decreases
with the system size following power laws, with exponents
$\gamma=0.87(2)$ for procedure (i), and $\gamma=0.24(2)$ for procedure
(ii) \cite{nota}. So, we can infer that the system seems
thermalised.

In Fig. \ref{fig:qft} we summarize the statistics of work during
non-equilibrium processes. We perform a forward process, $\alpha_i=0.2
\rightarrow \alpha_f=0.5$, and the corresponding backwards one,
$\alpha_i=0.5 \rightarrow \alpha_f=0.2$. For the first one, we prepare
two different initial states, with $E/N=-0.6$, by means of
procedures (i) and (ii). For the second one, we prepare $50$ different
initial states, with $25$ different energies, from $E/N=-0.26$ to
$E/N=-0.14$, with steps $\Delta E/N=5 \cdot 10^{-3}$ \cite{note2}, by
means of procedures (i) and (ii). All the processes are performed
following a TPM scheme \cite{Supl}. The quotient $P_f (E, \alpha_i;
w)/P_b(E+w,\alpha_f,-w)$ is displayed in panel (a) of
Fig. \ref{fig:qft} (see caption for details). Procedure (i) gives rise
to statistics compatible with Eq. (\ref{eq:qft}), but fluctuations of
work coming from procedure (ii) are totally different from the
expected. In panel (b) we show how the distance between numerics and
Eq. (\ref{eq:qft}) scales with the system size. Whereas this
distance decresases following a power law for procedure (i),
$N^{-\gamma}$ with $\gamma=0.78(6)$, we see no such decreasing for
procedure (ii); the relative error, $D$, in the last case is quite
large, around $10\%$, for system sizes between $N=10^3$ and $N=1.6
\cdot 10^4$. It is worth to remark that the region in which this error
is largest, $w=0.34-0.38$, is precisely the one that seemed perfectly
thermalised in Fig. \ref{fig:termo}.

\begin{figure}[h!]
  \includegraphics[width=0.4\linewidth,angle=-90]{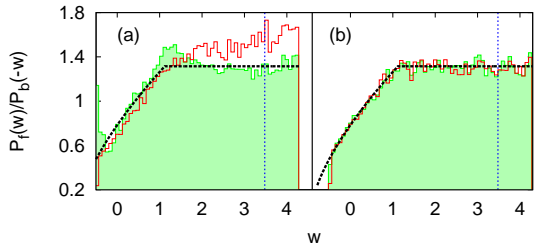} 
  \caption{Statistics of work of the procedures $\alpha=1.2$
    $\rightarrow$ $\alpha=2.0$ $\rightarrow$ $\alpha=0.6$ (empty
    histograms, red online), and $\alpha=1.2$ $\rightarrow$
    $\alpha=0.0$ $\rightarrow$ $\alpha=0.6$ (full histograms, green
    online), in the Dicke model. {\em Panel (a)}. Fock initial
    states. {\em Panel (b)} Microcanonical initial ensembles. The
    dotted black line represents the theoretical result from
    Eq. (\ref{eq:qft}). Vertical dotted line (blue online) indicate
    the cases displayed on panel (b) of Fig. \ref{fig:distr}.}
\label{fig:dicke}
\end{figure}

{\em Different trajectories.-} Here, we rely on the Dicke model
\cite{Dicke:54}, experimentally realized in \cite{Baumann:10,dicke_exp}, to test
how the statistics of work depend on the trajectory, if
non-equilibrium processes start from actual equilibrium states,
$\overline{\rho}$. The Dicke Hamiltonian models a system of $N$ two-level
atoms in a monochromatic radiation field,
\begin{equation}
H = \omega_o J_z + \omega a^{\dagger} a + \frac{\alpha}{\sqrt{N}} \left( J_+ + J_- \right) \left( a^{\dagger} + a \right).
\end{equation}
$\vec{J}$ is the pseudo-spin representation of $N$ two-level atoms,
with $N=2j$ (conserved). $a^{\dagger}$ ($a$) creates (annihilates) a
photon with frequency $\omega$. In all the calculations, $N=50$,
$\omega=\omega_o=1$, and the maximum number of photons is
$n_{\text{max}}=700$.

The Dicke model is known to be chaotic for large values of the
coupling constant $\alpha$ and energies above the ground-state region
\cite{Bastarrachea:16c,Relano:16,Buijsman:17}. In the thermodynamical
limit it is described by a classical Hamiltonian with two degrees of
freedom,
\begin{equation}
H = \omega_o j_z + \frac{\omega}{2} \left( q^2 + p^2 \right) + 2 \alpha \sqrt{j} \sqrt{1 - \frac{j_z^2}{j^2}} \cos \phi,
\end{equation}
allowing to obtain the density of states, $g \left(E, \alpha
\right)$, as we have done with the LMG model.

In this case, the consequences of Eq. (\ref{eq:condition}) are tested
by means of two different protocols, both starting from and ending at
the same values of the coupling constant: {\em (i)} $\alpha=1.2$
$\rightarrow$ $\alpha=2.0$ $\rightarrow$ $\alpha=0.6$, and {\em (ii)}
$\alpha=1.2$ $\rightarrow$ $\alpha=0.0$ $\rightarrow$
$\alpha=0.6$. Actual equilibrium states are obtained from initial Fock
states, $\left| n, m_j \right>$, giving rise to an energy $E=\omega_0
m_j + \omega n$; between all of them, we choose the best thermalised
one, for each energy. Thermalisation is tested by means of $J_z$,
$J_x^2$, $a^{\dagger} a$, and $\left(a^{\dagger} + a \right)^2$; we
compare the exact long-time average with the quantum microcanonical
average over a set of $51$ consecutive energy levels around the actual
energy $E$ \cite{nota3}. The average relative error for the four
observables and all the cases used to test Eq. (\ref{eq:qft}) (see
below for details) is $1.6 \cdot 10^{-2}$. A measure of chaos is also
performed. The average of $r_n = \text{min} \left( s_n/s_{n-1},
  s_{n-1}/s_n \right)$, where $s_n=E_{n-1} - E_n$, is obtained within
a window of $200$ levels around $E/j=-0.12$, for the case with
$\alpha=1.2$, $\left<r\right>=0.515(19)$. For the case with
$\alpha=0.6$, $\left<r \right> = 0.536(6)$ for the whole region $-0.6
\leq E/j \leq 4.2$. Since $\left<r\right>=0.5307(1)$ for ergodic
quantum systems, and $\left<r\right> = 2 \text{ln}2-1 \sim 0.386$ for
integrable ones \cite{Atas:13}, we can safely conlude that all our
numerical experiments are done within the chaotic region \cite{Supl}.

In panel (a) of Fig. \ref{fig:dicke} we summarize the statistics of
work of both procedures (see caption for details). The initial energy
for the forward process is $E/j=-0.12$; for the backward, we prepare
$60$ different initial states, with $-0.6 \leq E/j \leq 4.2$, with $\Delta
E/j =0.08$. In panel (b) we show the same calculation, but starting
from the microcanonical ensembles composed by $51$ consecutive levels
around the target energy, the same used to test
thermalisation. Results in panel (a) show that statistics of work
clearly depend on the trajectory, contrary to what states the
microcanonical quantum Crook's theorem; in particular, the first
one gives poor results for large values of the work. On the
contrary, panel (b) clearly show that both trajectories are totally
equivalent if the initial states are narrow microcanonical
ensembles. For the microcanonical initial states (obtained for $w \in
(0,4.3)$, to avoid data with few statistics), the average relative
errors form Eq. (\ref{eq:qft}) are $3.6 \%$, for trajectory (i), and
$3.2 \%$, for trajectory (ii). The same errors are $12.3 \%$ and $8.0
\%$, for Fock initial states.

\begin{figure}[h!]
  \includegraphics[width=0.6\linewidth,angle=-90]{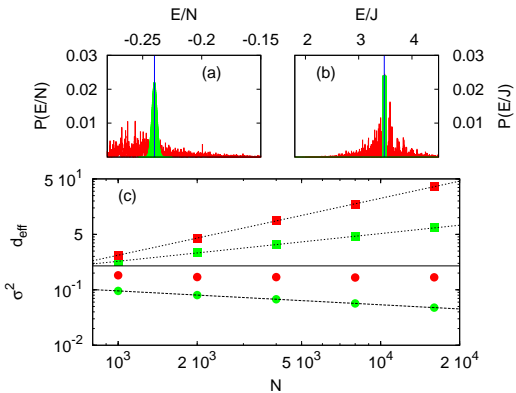}
  \caption{{\em Panel (a)}. Energy distributions resulting from
    procedure (i), filled curve (green online), and (ii), empty curve
    (red online); both are obtained with the LMG model with
    $\alpha=0.5$, $E/N=-0.24$ and $N=1.6 \cdot 10^4$. Dotted black
    line display the corresponding energies. {\em Panel (b)}. Energy
    distribution for the Dicke model with $\alpha=0.6$ and $E/j=3.48$:
    empty curve (red online) correspond to initial Fock state; filled
    curve (green online), to microcanonical ensemble. Again, dotted
    black line display the corresponding energies. {\em Panel
      (c)}. Scaling of the variance of the energy distribution,
    $\sigma^2$, as a function of the number of atoms, $N$ (solid
    circles); and scaling of the effective dimension of the Hilbert
    space, $d_{\text{eff}}$ (solid squares). Both are obtained with
    the LMG model. Light symbols (green online), results from
    procedure (i); dark symbols (red online), from procedure
    (ii). Dashed black line, power-law fit, $N^{-\gamma}$; dotted
    black line, power-law fit, $N^{\beta}$.}
\label{fig:distr}
\end{figure}

{\em Discussion.-} Fig. \ref{fig:distr} provides a deeper insight on
these results. Panels (a) and (b) show the energy distributions for
the cases enhanced in Figs. \ref{fig:qft} and \ref{fig:dicke} (see
caption for details). Panel (a) shows the results for the LMG
model. Procedure (i) gives rise to a narrow and smooth energy
distribution, whereas the one corresponding to procedure (ii) is wide
and erratic. Panel (b) displays the results for the Dicke model. The
initial Fock state gives rise to a distribution similar to the one
corresponding to procedure (ii) in panel (a), whereas the
microcanonical ensemble is pretty similar to the corresponding to
procedure (i). A quantitative test is performed in the LMG model, and
shown in panel (c). The variance of the distribution decreases
following a power law, $\sigma \sim N^{-\gamma}$, $\gamma=0.25003(3)$,
if the state is prepared with procedure (i), as it is expected
\cite{Schliemann:15}, but it remains approximately constant with
procedure (ii). A measure of the number of populated levels,
$d_{\text{eff}} = 1/\sum_n p_n^2$, where $p_n$ is the probability of
occupation of the level with energy $E_n$, is also displayed in the
Figure (see caption for details). In both cases, $d_{\text{eff}} \sim
N^{\beta}$, with $\beta=0.50013(3)$ for procedure (i), again as it is
expected \cite{Bastarrachea:16b}, and $\beta=1.027(7)$, for procedure
(ii). In both cases, $d_{\text{eff}}$ grows with the number of atoms,
as it is required for equilibration \cite{Gogolin:16}, but only in the
case of procedure (i) $d_{\text{eff}}$ represents a negligible part of
the spectrum, $d_{\text{eff}}/N \rightarrow 0$; in the case of
procedure (ii) the ratio $d_{\text{eff}}/N$ is approximately the same
for very different system sizes.

All these findings show that both procedure (ii) for the LMG model,
and Fock initial states for the Dicke model, produce initial states
too wide to fulfill the condition for the microcanonical quantum
Crook's theorem, Eq. (\ref{eq:condition}), but narrow enough to seem
properly thermalised. So, we conclude that

{\em The microcanonical quantum Crook's theorem consititues a more
  stringent test of thermalisation than the expected values of
  observables in equilibrium.}

Hence, experiments involving quantum fluctuation theorems
\cite{Chiara:15,An:15} can disclose non-proper thermalisation.

\begin{acknowledgments}
  This work has been supported by the Spanish Grant
  No. FIS2015-63770-P (MINECO/ FEDER). The author aknowledges
  A. L. Corps for his valuable comments.
\end{acknowledgments}


\begin{thebibliography}{99}
\bibitem{nota1} There is some controversy about the definition of work
  and the way to meassure it, in a non-equilibrium protocol (see, for
  example, \cite{Perarnau:17}). Here, we assume that a two-projective
  measurement (TPM) scheme is used. All our calculations are fully compatible with this scheme.
\bibitem{nota} We find power laws instead of exponential decays due to the
collective nature of the system, making the dimension of the Hilbert
space grow linearly (instead of exponentially) with the number of
atoms. Anyhow, both distances are small even at moderate system sizes,
$N \sim 1000$.
\bibitem{note2} Note that Eq. (\ref{eq:qft}) requires that the
  backward process starts at the energy resulting from the forward. If
  initial energy is $E$, and the measured work, $w$, the backward
  process must start from a thermalised state with energy $E + w$ to
  properly test Eq. (\ref{eq:qft}). Hence, we require just an initial
  state with energy $E$ for the forward process, but a number of
  initial states with energies $E + w_i$, covering all the measured
  values for the work, $w_i$.
\bibitem{nota3} In this case, we do not rely on a classical
  microcanonical average, as we have done with the LMG model, because
  of the finite-size effects related to the small number of atoms,
  $N=50$, accesible to exact diagonalization.
\end{thebibliography}
\end{document}